\documentclass[prb,aps,twocolumn,showpacs,superscriptaddress,floatfix,amsmath,amssymb]{revtex4}
\usepackage{graphicx}
\begin{document}

\title{Number Statistics of Ultracold Bosons in Optical Lattice}

\author{Yu Chuan Wen }
\affiliation{Interdisciplinary Center of Theoretical Studies, CAS,
Beijing 100080, China}
\author{Jing Yu Gan}
\affiliation{Center for Advanced Study, Tsinghua University,
Beijing, 100084, China}
\author{Xiancong Lu}
\affiliation{Institute of Theoretical Physics, CAS, Beijing
100080, China}
\author{Yue Yu}
\affiliation{Institute of Theoretical Physics, CAS, Beijing
100080, China}

\date{\today}

\begin{abstract}
We study the number statistics of ultracold bosons in optical
Lattice using the slave particle technique and quantum Monte Carlo
simulations. For homogeneous Bose-Hubbard model, we use the slave
particle technique to obtain the number statistics near the
superfluid to normal-liquid phase transition. The qualitatively
behavior agree with the recent experiment probing number fluctuation
[Phys. Rev. Lett. \textbf{96}, 090401 (2006)]. We also perform
quantum Monte Carlo simulations to 1D system with external harmonic
trap. The results qualitatively agree with the experiments.
\end{abstract}


\maketitle

\section{Introduction}\label{1}

The ultracold atoms in optical lattices have opened a new windows
to investigate the strongly correlated systems with highly tunable
parameters \cite{Bloch}. The basic physics of these ultracold
atoms is captured by the Bose Hubbard model, whose most
fundamental feature is the existence of superfluid to
Mott-insulator phase transition at zero temperature
\cite{fisher,Jaksch}. In a very shallow optical lattice, the
ultracold bosons are in superfluid phase and can be well described
by a macroscopic wave function with long-range phase coherence
\cite{gre}. In this case, the phase fluctuation is zero and the
on-site number fluctuation is large. When the optical lattice is
very deep, the bosons enter the Mott-insulating phase with fixed
number of atoms per site and without phase coherence, i.e., the
on-site number fluctuation is zero and the phase fluctuation is
large \cite{gre,Orzel}. The physics of the MI phase is that, when
the repulsive interaction between the atoms is large enough, the
number fluctuation would become energetically unfavorable and the
system would be in a number-squeezed state. This interaction
induced MI phase plays an important role in the strongly
correlated systems, as well as in various quantum information
processing schemes \cite{Rabl}.

In the past, some ultracold-atom experiments have been performed
to detected these number-squeezed MI phase through the observation
of increased phase fluctuations \cite{Stoferle,gre,Orzel} or
through an increased time scale for phase diffusion
\cite{Greiner}. Recently, the continuous suppression of on-site
number fluctuations was directly observed by Fabrice Gerbier et al
by monitoring the suppression of spin-changing collisions across
the superfluid/Mott-insulator transition \cite{Gerbier}. By using
a far off-resonant microwave field, the spin oscillations for
doubly occupied sites can be tuned into resonance and the
amplitude of spin oscillation is directly related to the
probability of finding atom pairs per lattice site. It was shown
by Fabrice Gerbier et al that, for small atom number, the
oscillation amplitude is increasingly suppressed with increasing
lattice depths and completely vanishes for large lattice depths.
In the MI region, this suppression persists up to some threshold
atom number. The authors also compared their experimental results
with the prediction of the Bose Hubbard model within a mean-field
approximation at zero temperature.

In this paper, we try to use the other approaches to study the
number fluctuation beyond the zero-temperature mean-field theory.
We first use the slave particle technique to obtained the number
statistics at the critical points of the superfluid to normal
liquid phase transition. The qualitative behaviors are the same as
the recent experiment \cite{Gerbier}. In the second part of this
paper, we step out the mean-field theory and perform quantum Monte
Carlo simulation to the 1D ultracold bosons with external harmonic
trap. The numerical results reproduce the qualitative behaviors of
the experiment.

This paper was organized as follows. In Sec. \ref{2}, we will
describe the slave particle technique to the homogeneous
Bose-Hubbard model. In Sec. \ref{3}, we will perform quantum Monte
Carlo simulation to the 1D ultracold bosons with external harmonic
trap. In Sec. \ref{4}, we will give our conclusions.

\section{Slave-particle approach to the number fluctuation of homogenous Bose-Hubbard model}\label{2}
We consider an ultracold atomic gas trapped in an
three-dimensional optical lattice potential,
$V_0(\textbf{r})=V_0\sum_{j=1}^3\sin^2(kr_j)$, with wave vectors
$k=2\pi/\lambda$ and $\lambda$ the laser wavelength. In real
experiments, an additional harmonic potential is superimposed to
the lattice potential; however, we only pay attention to the
homogeneous case in this section, which can be described by the
following homogeneous Bose-Hubbard Hamiltonian \cite{Jaksch},
\begin{eqnarray}\label{bhmodel}
H=-t\sum_{<ij>}a^\dag_ia_j-\mu\sum_in_i+\frac{U}{2}\sum_in_i(n_i-1).
\end{eqnarray}
Here $a_i^\dag$ is the creation operator at site $i$,
$n_i=a_i^\dag{a_i}$ is the particle number operator, and
$\langle{ij}\rangle$ denotes the sum over nearest neighbor sites.
$t$ and $U$ are the hopping amplitude and on-site interaction,
respectively,
\begin{eqnarray}
t&=&\int{d\textbf{r}w^*(\textbf{r}-\textbf{r}_i)\left(-\frac{\hbar^2}{2m}\nabla^2+
V_0(\textbf{r})\right)w(\textbf{r}-\textbf{r}_j)},\nonumber\\
U&=&g\int{d\textbf{r}|w(\textbf{r})|^4}.
\end{eqnarray}
In the following, we will use the slave particle technique to
obtain the finite temperature number fluctuation at the critical
points. In the slave particle language \cite{D,yy}, the bosonic
creation operator $a_{i}^{\dag}$ and annihilation operator $a_{i}$
can be decomposed into
\begin{eqnarray}
&&a_i^\dag=\sum_{\alpha=0}\sqrt{\alpha+1}|\alpha+1\rangle_{ii}\langle\alpha|,\nonumber\\
&&a_i=\sum_{\alpha=0}\sqrt{\alpha+1}|\alpha\rangle_{ii}\langle\alpha+1|,
\end{eqnarray}
where $|\alpha\rangle_{i}$ is the eigenstate of the particle
number operator $n_{i}=a_{i}^{\dag}a_{i}$ with $\alpha$ the
eigenvalue. In the slave particle language, every occupation state
is identified as a type of slave particle, i.e.,
$|\alpha\rangle_i$ and $_i\langle\alpha|$ are mapped to
$a_{\alpha,i}^{\dag}$ and $a_{\alpha,i}$, which are the slave
particle creation and annihilation operators, respectively. Then
$a_{i}^{\dag}$ and $a_{i}$ can be rewritten as
\begin{eqnarray}\label{transformation}
&&a_{i}^{\dag}=\sum_{\alpha=0}\sqrt{\alpha+1}a_{\alpha+1,i}^{\dag}a_{\alpha,i},\nonumber\\
&&a_i=\sum_{\alpha=0}\sqrt{\alpha+1}a_{\alpha,i}^{\dag}a_{\alpha+1,i}.
\end{eqnarray}
The slave particle operators are defined to satisfy the
anticommutation relation
$\{a_{\alpha,i},a_{\beta,j}^\dag\}=\delta_{\alpha\beta}\delta_{ij}$
in a slave fermion approach, and to satisfy the commutation
relation
$[a_{\alpha,i},a_{\beta,j}^\dag]=\delta_{\alpha\beta}\delta_{ij}$
in a slave boson approach. In order to reproduce the original
bosonic commutation relation $[a_{i},a_{j}^{\dag}]=\delta_{ij}$,
the slave particle operators must obey the constraint:
\begin{eqnarray}\label{constraint}
\sum_{\alpha=0}n_i^{\alpha}=\sum_{\alpha=0}a_{\alpha,i}^{\dag}a_{\alpha,i}=1.
\end{eqnarray}

Substituting the slave particle transformation
(\ref{transformation}) into Eq. (\ref{bhmodel}), the Bose-Hubbard
Hamiltonian can be replaced by
\begin{eqnarray}
H=&-&t\sum_{<ij>}\sum_{\alpha,\beta}\sqrt{\alpha+1}\sqrt{\beta+1}
a_{\alpha+1,i}^{\dag}a_{\alpha,i}a_{\beta,j}^{\dag}a_{\beta+1,j}\nonumber\\
&-&\mu\sum_i\sum_{\alpha}{\alpha}n_i^{\alpha}+\frac{U}{2}
\sum_i\sum_{\alpha}\alpha(\alpha-1)n_i^\alpha.
\end{eqnarray}
Following the steps in Refs. \cite{negele,stoof}, we write the
partition function as an imaginary time coherent state path
integral \cite{D,yy}:
\begin{eqnarray}
Z={\rm
Tr}e^{-{\beta}H}=\int{Da_{\alpha}D\bar{a}_{\alpha}D{\lambda}
e^{-S[\bar{a}_{\alpha},a_{\alpha},\lambda]}},
\end{eqnarray}
\begin{equation}
\begin{split}
S[\bar{a}_{\alpha},&a_{\alpha},\lambda]=\int_0^{\beta}{d\tau}\biggl[
\sum_i\sum_{\alpha}\bar{a}_{\alpha,i}
\biggr(\partial_{\tau}-\alpha\mu+\frac{U}{2}\alpha(\alpha-1)\\
&-i\lambda_i\biggr)a_{\alpha,i}+i\sum_i\lambda_i\\
&-t\sum_{<ij>}\sum_{\alpha,\beta}\sqrt{\alpha+1}\sqrt{\beta+1}
\bar{a}_{\alpha+1,i}a_{\alpha,i}\bar{a}_{\beta,j}a_{\beta+1,j}\biggr],
\end{split}
\end{equation}
where $\bar{a}_{\alpha,i}$ and $a_{\alpha,i}$ are introduced as
ordinary complex numbers in the slave boson approach, and as
Grassmann variables in the slave fermion approach satisfying the
Grassmann algebra. The Lagrange multiplier field $\lambda_i(\tau)$
comes from the constraint (\ref{constraint}), namely,
$\prod_i\delta(\sum_{\alpha}n_i^{\alpha}-1)$. The unit has been
set to $\hbar=k_B=1$ in all formulas. In order to decouple the
hopping term, we perform a Hubbard-Stratonovich transformation
\begin{eqnarray}
\lefteqn{\int{D}\Phi^*D\Phi\exp\left[-\int d\tau
t\sum_{<ij>}(\Phi_i^*-\sum_{\alpha}\sqrt{\alpha+1}
\bar{a}_{\alpha+1,i}a_{\alpha,i})\right.}\hspace{1.7cm}\nonumber\\
&&\left.\times(\Phi_j-\sum_{\alpha}\sqrt{\alpha+1}
\bar{a}_{\alpha,j}a_{\alpha+1,j})\right].
\end{eqnarray}
The Hubbard-Stratonovich field $\Phi_i$ introduced here can be
identified as the order parameter of superfluid for
$\langle\Phi_i\rangle=\langle\sum_{\alpha}\sqrt{\alpha+1}
\bar{a}_{\alpha,i}a_{\alpha+1,i})\rangle =\langle{a_i}\rangle$. We
then perform a Fourier transform on all the fields $A_i$ by
\begin{eqnarray}
A_i=\frac{1}{\sqrt{L\beta}}\sum_{\textbf{k},n}A_{\textbf{k}n}e^{i(\textbf{k}
\cdot{\textbf{r}_i}-\omega_n\tau)},
\end{eqnarray}
where $L$ is the total number of sites of the optical lattice and
$\omega_n$ is the Matsubara frequency, which equals
$(2n+1)\pi/\beta$ or $2n\pi/\beta$ for fermionic or bosonic
fields. After relaxing the constraint (\ref{constraint}) to one
slave particle per site on average over the whole lattice, i.e.,
replacing $\lambda_{\textbf{k},n}$ with a constant
$\lambda_{\textbf{0},0}$, we arrive at an effective action divided
into two parts,
\begin{eqnarray}
S_{eff}[\Phi,a_{\alpha},\lambda]=S_0+S_I,
\end{eqnarray}
\begin{eqnarray*}
S_0=iL\beta\lambda&+&\sum_{\textbf{k},n}\sum_{\alpha}
\bar{a}^\alpha_{\textbf{k},n}\left[-i\omega_n
+c(\alpha)\right]a^\alpha_{\textbf{k},n}\nonumber\\
&+&\sum_{\textbf{k},n}\epsilon_{\textbf{k}}|\Phi_{\textbf{k},n}|^2=S_0^{sp}
+\sum_{\textbf{k},n}\epsilon_{\textbf{k}}|\Phi_{\textbf{k},n}|^2,
\end{eqnarray*}
\begin{eqnarray*}
S_I&=&-\sum_{\textbf{k},\textbf{k}',n,n'}\sum_{\alpha}
\frac{\epsilon_{\textbf{k}'}}{\sqrt{L\beta}}\left[\left(\sqrt{\alpha+1}
\bar{a}^{\alpha+1}_{(\textbf{k+k}'),(n+n')}
a^\alpha_{\textbf{k},n}\right)\right.\\
&&\left.\times\Phi_{{\textbf{k}',n'}}+{\left(\sqrt{\alpha+1}
\bar{a}^\alpha_{\textbf{k},n}a^{\alpha+1}_{(\textbf{k+k}'),(n+n')}
\right)\Phi_{{\textbf{k}',n'}}^*}\right],
\end{eqnarray*}
where $c(\alpha)=-i\lambda-\alpha\mu+\alpha(\alpha-1)U/2$,
$\lambda=\lambda_{\textbf{0},0}/\sqrt{L\beta}$, and
$\epsilon_{\textbf{k}}=2t\sum_{i=1}^d\cos({k_i}a)$ with $d$ and
$a$ being the dimension and spacing constant of the lattice.

Near the critical point, the order parameter $\Phi$ is small and
the perturbation can be performed in terms of $S_I$. The partition
function $Z_0$ of non-interacting slave particles comes from the
contribution of the zeroth-order term $S^{sp}_0$ which is given by
\begin{eqnarray}\label{z0}
Z_0=e^{-\beta\Omega_0}=\int{Da_{\alpha}D\bar{a}_{\alpha}e^{-S_0^{sp}}}.
\end{eqnarray}
Here $\Omega_0$ is the zeroth-order thermodynamic potential,
\begin{eqnarray}\label{omega0}
-\Omega_0=i\lambda{L}\pm\frac{L}{\beta}\sum_{\alpha}\ln(1\pm
e^{-\beta c(\alpha)}),
\end{eqnarray}
where the $+(-)$ sign corresponds to the slave fermion (slave
boson). After expanding $e^{-S_{eff}}$ up to second order of $S_I$
and integrating out the slave particle field, we obtain the new
effective action to second order of the order parameter field,
\begin{equation}\label{e,eff2}
S_{E,eff}[\Phi^*,\Phi]=\beta\Omega_0
-\sum_{\textbf{k},n}\Phi^*_{\textbf{k},n}G^{-1}(\textbf{k},i\omega_n)\Phi_{\textbf{k},n}.
\end{equation}
The Green's function $G(\textbf{k},i\omega_n)$ is defined by
\begin{equation}\label{green}
-G^{-1}(\textbf{k},i\omega_n)=\epsilon_{\textbf{k}}-\epsilon_{\textbf{k}}^2\sum_{\alpha}
(\alpha+1)\frac{n^{\alpha}-n^{\alpha+1}}{-i\omega_n-\mu+\alpha{U}},
\end{equation}
where $n^{\alpha}$ is the occupation number and equal to
\begin{equation}\label{occupation number}
n^{\alpha}=\frac{1}{\exp\{\beta[-i\lambda-\alpha\mu+\alpha(\alpha-1)U/2]\}\pm1},
\end{equation}
in which the $+$ and $-$ sign correspond to slave fermion and
slave boson, respectively.

The saddle point approximation to the constraint field $\lambda$
means: $\partial\Omega/\partial\lambda=0$; the particle number
conservation condition requires $-\partial\Omega/\partial\mu=N$.
The mean-field approximation means all the fluctuations coming
from the Green's function would not be considered in the above two
conditions. Then the following two mean-field equations can be
derived,
\begin{eqnarray}
&&\sum_{\alpha=0}n^{\alpha}=1,\\\label{constrain}
&&\sum_{\alpha=0}{\alpha}n^{\alpha}=\frac{N}{L}=n,
\end{eqnarray}
where the $n=N/L$ is the average particle density.
\begin{figure}
\begin{center}
\includegraphics[width=0.9\columnwidth]{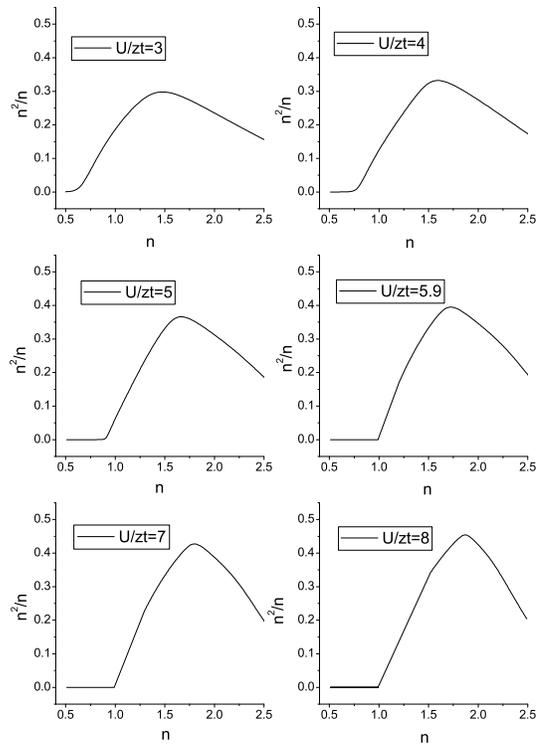}\\
\caption{\label{fig1} The average probability of a site to be
occupied by two bosons $n^2/n$ as a function of average particle
density $n$, for different interactions $U/zt$. All the
calculations are performed at the critical temperature of the
phase transition.}
\end{center}
\end{figure}

Combining Eqs. (\ref{green})-(\ref{constrain}), we can obtain the
number fluctuation at critical temperature of superfluid to normal
liquid phase transition. We show the results in Fig.\ref{fig1}.
One can see that the qualitative behavior agree with the recent
experiment probing number fluctuation [Phys. Rev. Lett.
\textbf{96}, 090401 (2006)].

\section{NUMERICAl RESULTS: QUANTUM MONTE CARLO CALCULATION IN ONE DIMENSION}\label{3}
\begin{figure}[t]
\includegraphics[width=1\columnwidth]{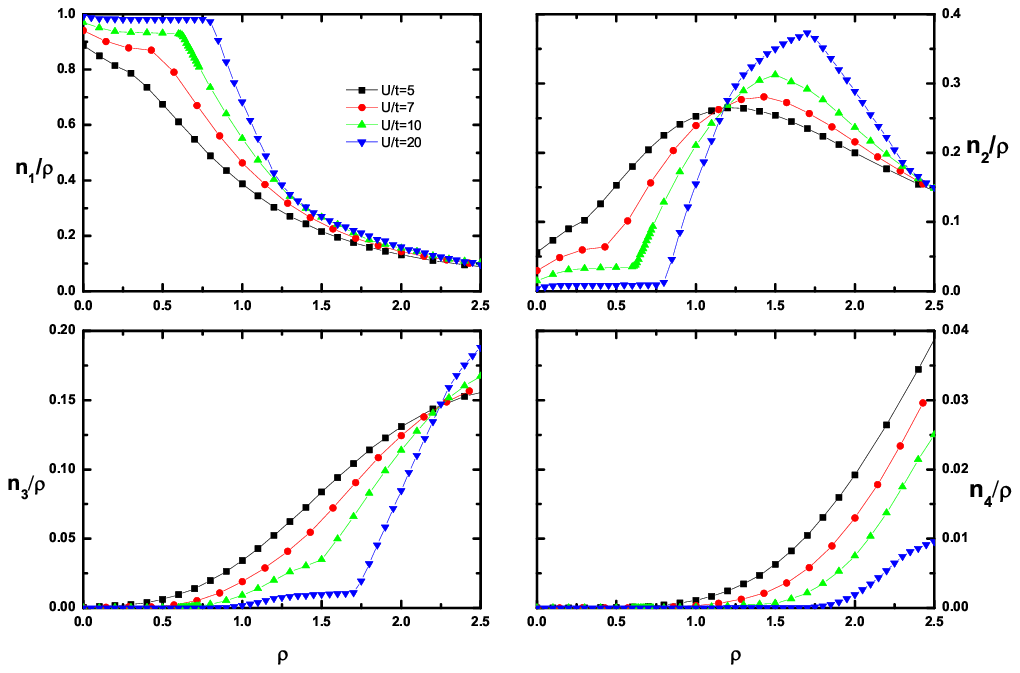}
\caption{(Color on line)  $n_\alpha/\rho$ as a function of boson
density $\rho$ for various $U/t$ at $V_t=0.01t$. The chain size is
$L=100$. $n_\alpha(\alpha=1,2,3,4)$ are average bonson density for
$\alpha$ bosons on each site.}\label{n1n2}
\end{figure}

\begin{figure}[t]
\includegraphics[width=8cm]{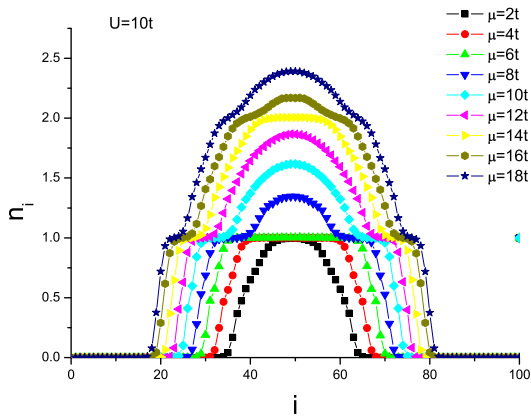}
\caption{(Color on line)  Local particle density $n_i$ as a
function of site $i$ for various $\mu$ at $U=10t$.}\label{rho}
\end{figure}
\begin{figure}[t]
\includegraphics[width=1\columnwidth]{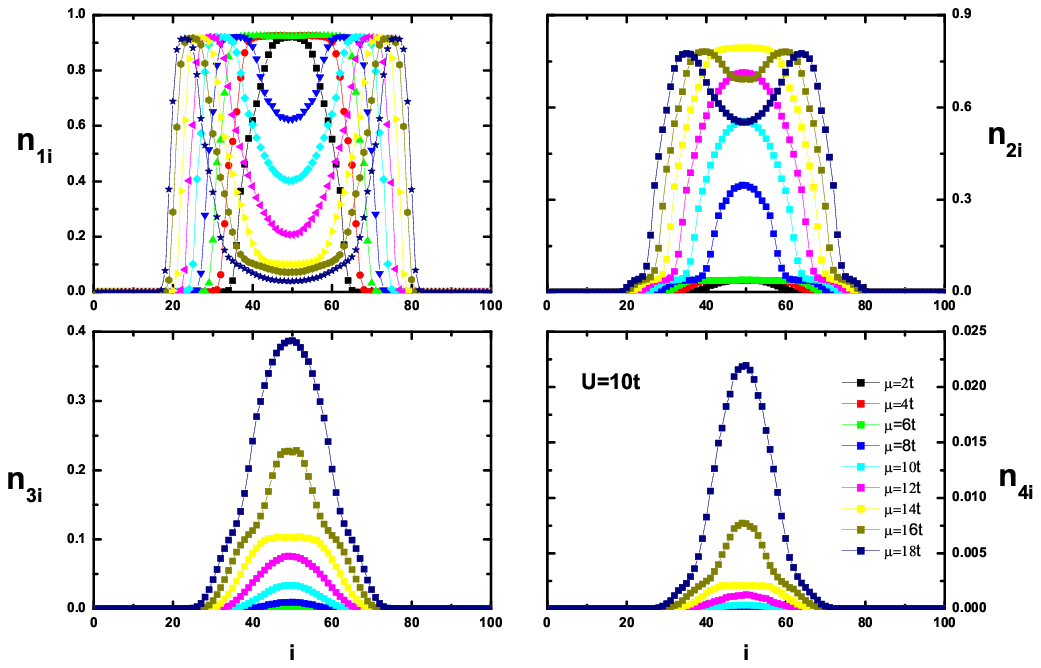}
\caption{(Color on line)  $n_\alpha/\rho$ as a function of boson
density $\rho$ for various $U/t$ at $V_t=0.01t$. The chain size is
$L=100$. $n_\alpha(\alpha=1,2,3,4)$ are average bonson density for
$\alpha$ bosons on each site.}\label{onsite}
\end{figure}

Although mean field theory are applicable in higher dimensions,
its application to 1D is questionable due to the large quantum
fluctuations. In this section, we will come to the numerical
calculation and focus our study on the Boson$-$Hubbard model (1)
in a harmonic trapped potential in 1D optical lattice. The trapped
potential we use is:
\begin{eqnarray}
V_T &= & V_t\sum_i (i-L/2)^2
\end{eqnarray}
where $L$ is the chain length and $V_t=0.02t$. The method we use
is quantum Monte Carlo (QMC) simulations using the stochastic
series expansion technique~\cite{sandvik1,sandvik2}. In the
simulation, we set the lattice is large enough to neglect the
boundary effects and the inverse of temperature $\beta=100t$ in
order to reach the ground state properties.

Fig.\ref{n1n2} show $n_\alpha/\rho$($\alpha=1,2,3,4$) v.s. the
average boson density $\rho$, where $n_\alpha$ is the average
boson density for $\alpha$ bosons of the system. From the figure
we can see, $n_1$ decreases with increasing average boson density
$\rho$, while $n_3$, $n_4$ increase with it. They all change
monotonously while $n_2$ is nonmonotonous. When $\rho$ is small,
$n_1$ is quite large, i.e., most of the sites are one particle
occupied. In this region, there is a $n=1$ Mott plateau in the
middle of the chain(see fig.\ref{rho}). As $\rho$ increases, $n_1$
decreases while $n_2$ increases slowly. At some critical
$\rho_{c1}$, $n_2$ suddenly increases fast. This is because when
$\rho>\rho_{c1}$, a superfluid forms in the middle of the $n=1$
Mott plateau in the middle of the chain. As $\rho$ increase
further, $n_2$ reaches its maximum at $\rho_{c2}$ and then
decreases with increasing $\rho$. This corresponds the formation
of the $n=2$ Mott plateau in the middle of the chain(also see
Fig.\ref{rho}). When $U$ increase, both $\rho_{c1}$ and
$\rho_{c2}$ increase. From the above discussion we can see that,
$n_2$ can be used to describe the Mott-superfluid transition. Our
results are qualitatively agree with Gerbier's experiment and the
mean field theory above.

In Fig.\ref{rho} and Fig.\ref{onsite}, we show the local particle
density and onsite number statistics for various $\mu$ at $U=10t$.
From Fig.\ref{rho}, we can see that the formation of the $n=1$ and
$n=2$ plateau with increasing $\mu$. In Fig.\ref{onsite} we find
that $n_1$ develop concaves in the center of the trap. And these
concaves become more and more deep which correspond with the
decreasing of the total number of $n_1$ in Fig.\ref{n1n2}. $n_2$
evolves nonmonotonously. It increase suddenly with the appearing
of the superfluid region in the center of the trap and develop
concave in the center of the trap with the appearing of the second
Mott plateau. With the increasing of boson number, the $n_3$ and
$n_4$ are not important when the total number of bosons are not
large.

\section{CONCLUSION}\label{4}

We studied the number fluctuation of ultracold bosons in optical
Lattice using the slave particle technique and quantum Monte Carlo
simulation. By using the slave particle technique, we obtained the
number statistic at the critical points of superfluid to normal
liquid phase transition. We also performed QMC simulations to the
one dimension Bose-Hubbard model with external harmonic trap. The
numerical results qualitatively agree with the recent experiment
\cite{Gerbier}.

\begin{acknowledgments}
This work was supported in part by Chinese National Natural
Science Foundation. The simulations were performed on the HP-SC45
Sigma-X parallel computer of ITP and ICTS, CAS .
\end{acknowledgments}

\textit{Note added.}-Recently we became aware of a parallel
numerical work \cite{prokof} that reach similar conclusions.

\end{document}